\documentclass[twocolumn,showpacs,preprintnumbers,amsmath,amssymb]{revtex4}
\usepackage{graphicx}% Include figure files
\usepackage{dcolumn}% Align table columns on decimal point
\usepackage{bm}% bold math

\begin{document}

\preprint{}

\title{Quantum-correlating power of local quantum channels}% Force line breaks with \\

\author{Xueyuan Hu}
\email{xyhu@iphy.ac.cn}
\author{Heng Fan}
\author{D. L. Zhou}
\author{Wu-Ming Liu}
\affiliation{Beijing National Laboratory for Condensed Matter
Physics, Institute of Physics, Chinese Academy of Sciences, Beijing
100190, China}
\date{\today}% It is always \today, today,

\begin{abstract}
Quantum correlation can be created by local operations from a
classically correlated state. We define quantum-correlating power
(QCP) of a local quantum channel as the maximum amount of quantum
correlation that can be created by the channel. The quantum
correlation that we discuss in this article is defined on the left
part of the bipartite state. We prove that for any local channel,
the optimal input state, which corresponds to the maximum quantum
correlation in the output state, must be a classical-classical
state. Further, the single-qubit channels with maximum QCP can be
found in the class of rank-1 channels which take their optimal input
states to rank-2 quantum-classical states. The analytic expression
for QCP of single-qubit amplitude damping channel is obtained.
Super-activation property of QCP, i.e., two zero-QCP channels can
consist a positive-QCP channel, is discussed for single-qubit phase
damping channels.
\end{abstract}

\maketitle

\section{Introduction}

The quantum nature of correlation goes beyond quantum entanglement.
There are separable states that contain correlation with no
classical counterpart. Researches show that such separable states
can also be useful for quantum computation
\cite{PhysRevLett.100.050502}, quantum state discrimination
\cite{PhysRevLett.107.080401} and quantum communication
\cite{arXiv:1203.1268v2,arXiv:1203.1629v1,arXiv:1203.1264v2}. The
importance of quantum correlation also lies in its close connection
to quantum entanglement
\cite{PhysRevLett.106.220403,PhysRevLett.106.160401,PhysRevLett.107.020502}.
Various ways for detecting and measuring the quantum correlation
have been proposed
\cite{PhysRevLett.88.017901,PhysRevLett.105.190502,PhysRevA.67.012320,PhysRevLett.106.120401,PhysRevLett.108.120502}
and the dynamics of quantum correlation under noises are also
studied, see for example \cite{PhysRevLett.104.200401}. Among the
different measures for quantum correlation, the quantum discord and
one-way quantum deficit receives their physical interpretations
\cite{RevModPhys.81.1,PhysRevA.83.032323,PhysRevA.83.032324}.

Counterintuitively, local operation can create quantum correlation
in some classically correlated states
\cite{PhysRevLett.107.170502,PhysRevA.85.032102,arXiv:1112.5700v1,PhysRevA.85.010102,PhysRevA.85.022108}.
In particular, any separable state with positive quantum discord can
be produced by local positive operator-valued measure (POVM) on a
classical state in a larger Hilbert space \cite{PhysRevA.78.024303}.
The criteria for checking whether a local trace-preserving operation
is able to generate quantum correlation has recently been obtained.
For a single-qubit channel, it can create quantum correlation in
some classical states if and only if it is neither a unital channel
nor a classical channel \cite{PhysRevLett.107.170502}. For a quantum
channel of arbitrary finite dimension, it is able to create quantum
correlation if and only if it is not a commutativity-preserving
channel \cite{PhysRevA.85.032102,arXiv:1112.5700v1}.

On solving the problem whether a local channel can create quantum
discord, it is natural to ask the following question: how much
quantum correlation can be built by local operation? In this
article, we investigate the problem by defining quantum-correlating
power of a local quantum channel, which quantifies the maximum
quantum correlation that can be generated by the channel. For any
local channels, the input state which corresponds to the maximum
quantum correlation in the output state is proved to be a
classical-classical state. Further, the quantum state with maximum
quantum correlation which is obtained local operation on a two-qubit
classical-quantum state can be found in the class of rank-2
quantum-classical states. The QCP of amplitude damping channel is
calculated as an example. The interesting effect that two zero-QCP
channels can consist a positive-QCP channel is observed, and is
named as the super-activation of QCP. As a by-product, we find a
class of states with zero pair-wise correlation but non-zero genuine
quantum correlation.

\section{Quantum-correlating power and optimal input state}

Generally, a state is said to have zero quantum correlation on $A$
if and only if there is a measurement on $A$ that does not affect
the total state. Such states are called classical-quantum states. We
label $\mathcal C_0$ as the set of all classical-quantum states.
Then $\mathcal C_0$ can be written as \cite{arXiv:0807.4490v1}
\begin{equation}
\mathcal C_0=\{\rho|\rho=\sum_iq_i\Pi_{\alpha_i}^A\otimes\rho_i^B\},
\end{equation}
where $\{\Pi_{\alpha_i}^A=|\alpha_i\rangle\langle\alpha_i|\}$ are a
set of orthogonal basis of part $A$.

Various measures for quantifying quantum correlation have been
proposed. For example, quantum discord \cite{PhysRevLett.88.017901}
is defined as the minimum part of the mutual information shared
between $A$ and $B$ that cannot be obtained by the measurement on
$A$
\begin{equation}
\delta_{B|A}(\rho)=\min_{\{F^A_i\}}S_{B|A}(\rho_{F^A_iB})-S_{B|A}(\rho),
\end{equation}
where $S_{A|B}(\rho)=S(\rho)-S(\rho_B)$ with
$S(\rho)=-\mathrm{Tr}(\rho\log_2\rho)$ is conditional entropy,
$\{F_i^A\}$ is a POVM on qudit $A$, and
$\rho_{F^A_iB}=\sum_iF^A_i\rho F^{A\dagger}_i$ is the state of
qudits $A$ and $B$ after the POVM. It has been proved that, for
separable states, the optimal POVM is just von Neumann measurement
$\{\Pi^A_i\}$ \cite{arXiv:0807.4490v1}. Another example is the
distance-based measure of quantum correlation
\cite{PhysRevLett.107.170502}
\begin{equation}
Q_D(\rho)=\min_{\sigma\in\mathcal C_0}D(\rho,\sigma),\label{QD}
\end{equation}
where the state distance satisfies the property that $D$ does not
increase under any quantum operation. Trace-norm distance
$D_1=\mathrm{Tr}|\rho-\sigma|/2$ with $|\hat O|=\sqrt{\hat
O^{\dagger}\hat O}$ and relative entropy
$S(\rho\parallel\sigma)=\mathrm{Tr}[\rho(\log_2\rho-\log_2\sigma)]$
are examples satisfying this property \cite{book1}. One-way quantum
deficit
\begin{equation}
\Delta_{B|A}^{\leftarrow}=\min_{\{\Pi_A\}}S(\rho_{\Pi^A_iB})-S(\rho),\label{deficit}
\end{equation}
is in fact the minimum relative entropy to classical-quantum states
\cite{PhysRevA.71.062307}, and thus belongs to this class of quantum
correlation measure. Notice that the measures of quantum correlation
are asymmetric for $A$ and $B$. Here and after, we only discuss the
quantum correlation defined on $A$.

The measures of quantum correlation $Q$ we discuss in this article
satisfy the following three conditions. (a) $Q(\rho)=0$ iff
$\rho\in\mathcal C_0$; (b) $Q(U\rho U^{\dagger})=Q(\rho)$ where $U$
is a local unitary operator on $A$ or $B$; (c)
$Q(I\otimes\Lambda_B(\rho))\leq Q(\rho)$. Conditions (a) and (b) are
satisfied by most of the quantum correlation measures. It has been
proved that quantum discord satisfies condition (c)
\cite{PhysRevLett.106.160401}. Here we briefly prove that $Q_D$
satisfies condition (c). Suppose the closest classical-quantum state
to $\rho$ is labeled as $\sigma$, then we have
$Q_D(\rho)=D(\rho,\sigma)\geq
D(\Lambda_B(\rho),\Lambda_B(\sigma))\geq Q_D(\Lambda_B(\rho))$. The
last inequation holds because $\Lambda_B(\sigma)$ is still a
quantum-classical state, but may not be the closest one to
$\Lambda_B(\rho)$. It should be noticed that geometric quantum
discord does not satisfy condition (c) \cite{footnote}, and is thus
out of the scope of this paper.

Local operations on $A$ can create quantum correlation from a
classical-quantum state. In order to characterize how much quantum
correlation can be created by a local channel, we introduce the
definition of quantum-correlating power.

\emph{Definition (quantum-correlating power).} The
quantum-correlating power of a quantum channel is defined as
\begin{equation}
\mathcal Q(\Lambda)=\max_{\rho\in\mathcal C_0}Q(\Lambda\otimes
I(\rho)),\label{QCP}
\end{equation}
where $Q$ is a measure of quantum correlation which satisfies
conditions (a-c).

The input state $\rho$ that corresponds to the maximization in Eq.
(\ref{QCP}) is called the optimal input state. Here we give a
general form of the optimal input state.

\emph{Theorem 1.} For any $d$-dimension local channel acting on $A$,
The optimal input state with the maximum amount of quantum
correlation in the output state is a classical-classical state of
form
\begin{equation}
\varrho=\sum_{j=0}^{d-1}q_j\Pi_{\alpha_j}^A\otimes\Pi_{\beta_j}^B,\label{optimal_input}
\end{equation}
where $\{\Pi_{\beta_j}^B=|\beta_j\rangle\langle\beta_j|\}$ is the
orthogonal basis for the Hilbert space of qudit $B$.

\emph{Proof.} Consider a classical-quantum state
$\varrho'\in\mathcal C_0$ as input state. After a local channel on
$A$, the state becomes
\begin{equation}
\rho'=\sum_iq_i\Lambda(\Pi_{\alpha_i}^A)\otimes\rho_i^B.\label{output_1}
\end{equation}
For input state $\varrho$ as in Eq. (\ref{optimal_input}), the
corresponding output state is
\begin{equation}
\rho=\sum_iq_i\Lambda(\Pi_{\alpha_i}^A)\otimes\Pi_{\beta_j}^B.\label{output_2}
\end{equation}
We first prove that $\rho'$ can be prepared from $\rho$ by a local
operation on $B$. Writing the $d$ states of qudit $B$ in Eq.
(\ref{output_1}) as
$\rho_k^B=\sum_{i=0}^{d-1}\lambda_i^{(k)}|\phi_i^{(k)}\rangle_B\langle\phi_i^{(k)}|,\
k=1,\cdots,d$, we find a rank-1 channel
$\Lambda_B(\cdot)=\sum_{i=0}^{d-1}\sum_{k=1}^d\lambda_i^{(k)}E_i^{(k)}(\cdot)E_i^{(k)\dagger}$,
where $E_i^{(k)}=|\phi_i^{(k)}\rangle\langle \beta_k|$, such that
$\Lambda_B(\Pi_{\beta_j}^B)=\rho_k^B$. It means that
$\rho=I\otimes\Lambda_B(\rho')$. Reminding that local operation on
$B$ never increase the quantum correlation on $A$, we have
$Q(\rho)\geq Q(\rho')$. It means that for any state $\rho'$ in the
form of Eq. (\ref{output_1}), we can always find a state $\rho$ in
the form of Eq. (\ref{output_2}), whose quantum correlation is
larger than $\rho'$. Therefore, state in form of Eq.
(\ref{optimal_input}) is the optimal input state. This completes the
proof of Theorem 1.

\section{Channels with maximum QCP}

We have investigated the maximum quantum correlation that can be
created by a given local channel. It is also interesting to ask the
following question: how much quantum correlation can be generated
from a classical state when all the local quantum operation is
allowed? In this section, we focus on finding the single-qubit
channels with maximum QCP.

\emph{Lemma 1.} For any two states of a qubit $\rho_j$, $j=0,1$,
there exist two pure states $|\psi\rangle$ and $|\phi\rangle$, such
that
$\rho_j=p_j|\phi\rangle\langle\phi|+(1-p_j)|\psi\rangle\langle\psi|$,
where $0\leq p_j\leq1$, $j=1,2$.

\emph{Proof.} We will first prove that for any two states of a qubit
$\rho_1$ and $\rho_2$, there exist a pure state $|\psi\rangle$, such
that
\begin{equation}
\rho_1=p\rho_2+(1-p)|\psi\rangle\langle\psi|, 0\leq
p\leq1.\label{decom1}
\end{equation}
We discuss this problem in the Bloch presentation:
$\rho_j=(I+\vec{c_j}\cdot\vec{\sigma})/2$, $j=1,2$, and
$|\psi\rangle\langle\psi|=(I+\vec{a}\cdot\vec{\sigma})/2$, where
$\vec{\sigma}=\{\sigma_x,\sigma_y,\sigma_z\}$ are Pauli matrices,
$\vec{c_j}=\mathrm{Tr}(\rho_j\vec{\sigma})$ and
$\vec{a}=\langle\psi|\vec{\sigma}|\psi\rangle$. Then Eq.
(\ref{decom1}) is equivalent to
\begin{equation}
\vec{c_1}=p\vec{c_2}+(1-p)\vec{a},\label{decom}
\end{equation}
where $|\vec{c_j}|\leq1$ and $0\leq p\leq1$. $|\vec{a}|=1$ leads to
$p=[1-\vec{c_1}\cdot\vec{c_2}-\sqrt{(1-\vec{c_1}\cdot\vec{c_2})^2-(1-|\vec{c_1}|^2)(1-|\vec{c_2}|^2)}]/(1-|\vec{c_2}|^2)$.
It is straight forward to verify that $0\leq p\leq1$. Therefore, Eq.
(\ref{decom}) holds.

Consequently, for $\rho_2$ and $|\psi\rangle$, we can always find a
pure state $|\phi\rangle$ such that
\begin{equation}
\rho_2=p'|\psi\rangle\langle\psi|+(1-p')|\phi\rangle\langle\phi|,
0\leq p'\leq1.\label{decom2}
\end{equation}
Combining Eqs. (\ref{decom1}) and (\ref{decom2}), we have
$\rho_1=(1-p+pp')|\psi\rangle\langle\psi|+p(1-p')|\phi\rangle\langle\phi|$.
This completes the proof of Lemma 1. It is worth mentioning that
$\vec{a}$ and $\vec{b}\equiv\langle\phi|\vec{\sigma}|\phi\rangle$
are just the two intersections of the Bloch sphere surface and the
line $\overline{c_0c_1}$, where $\overline{c_0c_1}$ is the line
fixed by the two points $\vec{c_0}$ and $\vec{c_1}$.

Now we are ready to prove the second central result of this paper.

\emph{Theorem 2.} The local single-qubit channel with maximum QCP
can be found in the set of channels
\begin{equation}
\mathcal
D_0=\{\Lambda|\Lambda(\cdot)=\sum_{i=0}^1E_i(\cdot)E_i^{\dagger},E_i=|\psi_i\rangle\langle\alpha_i|\},\label{optimal_channel}
\end{equation}
where $|\psi_0\rangle$ and $|\psi_1\rangle$ are two non-orthogonal
pure states.

\emph{Proof.} In order to find out the maximum-QCP channel, we
investigate the form of optimal output state, which contains the
maximum quantum correlation created by local operations on a
classical-classical state. The optimal output state can be found in
the subset of the rank-2 quantum-classical state
\begin{equation}
\tilde{\mathcal
C}_0\equiv\{\tilde{\rho}|\tilde{\rho}=\sum_{i=0}^{1}p_i|\psi_ii\rangle\langle\psi_ii|\}.\label{optimal_output}
\end{equation}
The reason is as follows. Consider the optimal input state as in Eq.
(\ref{optimal_input}) and the corresponding output state as in Eq.
(\ref{output_2}) with $d=2$. According to lemma 1, each
$\rho_j\equiv\Lambda(\Pi_j^A)$ can be decomposed as
$\rho_j=\sum_{i=0}^{d}p^{(j)}_i|\psi_i\rangle\langle\psi_i|$,
$j=0,1$, and consequently, Eq. (\ref{output_2}) can be written as
\begin{equation}
\rho=\sum_{i=0}^{1}p_i|\phi_i\rangle\langle\phi_i|\otimes\xi_i,\label{output1}
\end{equation}
where $p_i=\sum_{j=0}^{d-1}q_jp_i^{(j)}$ and
$\xi_i=(\sum_{j=0}^{d-1}q_jp_i^{(j)}|j\rangle\langle j|)/p_i$ for
$i=0,1$. From the proof of theorem 1, any state $\rho$ in form of
Eq. (\ref{output1}) can be obtained from $\tilde{\rho}$ in Eq.
(\ref{optimal_output}) by some local operations on $B$. Meanwhile,
the quantum correlation we discuss here can not be increased by
local operation on $B$. Therefore, the optimal output state which
contains the maximum correlation can be found in $\tilde{\mathcal
C}_0$. Further, for any output state $\tilde{\rho}\in\tilde{\mathcal
C}_0$, we can find a channel $\Lambda\in\mathcal D_0$ which takes a
classical input state to $\tilde{\rho}$. This completes the proof of
theorem 2.

Based on theorem 2, we derive the local single-qubit channel with
the maximum QCP based on quantum discord. We first need to find
$\tilde{\rho}=p_0|00\rangle\langle00|+p_1|\phi1\rangle\langle\phi1|$
in $\tilde{\mathcal C}_0$ which contains the maximum quantum
discord. The quantum discord of a rank-2 two-qubit state can be
calculated analytically using the Koashi-Winter relation
\cite{PhysRevA.69.022309}
\begin{equation}
\delta_{B|A}=\mathcal E_{BC}+S_{B|C},
\end{equation}
where $\mathcal E_{BC}$ is the entanglement of formation (EOF)
between qubits $B$ and $C$, and qubit $C$ is the purification of
state $\tilde{\rho}$
\begin{equation}
|\Psi\rangle_{ABC}=\sqrt{p_0}|000\rangle+\sqrt{p_1}|\phi11\rangle.
\end{equation}
Therefore, we have
\begin{equation}
\delta_{B|A}(\tilde{\rho})=h(\sqrt{1-t^2}\sin\phi)+h(\sqrt{1-(1-t^2)\sin^2\phi})-h(t),\label{max_QCP}
\end{equation}
where $t=p_0-p_1$. Eq. (\ref{max_QCP}) reaches its maximum
$\delta_{max}\approx0.2017$ at $\phi=\pi/4$ and $t=0$. Therefore,
the channels with maximum QCP should satisfy
$\Lambda^{Max}(|\phi\rangle\langle\phi|)=|\phi\rangle\langle\phi|$
and
$\Lambda^{Max}(|\phi+\pi/2\rangle\langle\phi+\pi/2|)=|\phi+3\pi/4\rangle\langle\phi+3\pi/4|$.
It is direct forward to write a class of maximum-QCP channels, which
are unitarily equivalent to
$\tilde{\Lambda}(\cdot)=\sum_{i=0}^1\tilde{E}_i(\cdot)\tilde{E}_i^{\dagger}$,
where
\begin{equation}
\tilde{E}_0=|0\rangle\langle0|,\tilde{E}_1=|+\rangle\langle1|,
\end{equation}
and the corresponding QCP is
\begin{equation}
\mathcal
Q_{\delta}(\Lambda^{Max})=2h(\frac{1}{\sqrt2})-1\approx0.2017.
\end{equation}

It is worth mentioning that there are separable states containing
larger quantum discord. For example, for separable state
$\rho=(|\Phi^+\rangle\langle\Phi^+|+|\Psi^+\rangle\langle\Psi^+|)/2$
with $|\Phi^+\rangle=(|00\rangle+|11\rangle)/\sqrt2$ and
$|\Psi^+\rangle=(|01\rangle+|10\rangle)/\sqrt2$ two Bell states, the
quantum discord is $\delta(\rho)=3/4$, according to the result of
Ref. \cite{PhysRevA.77.042303}. Such states can not be prepared by
local operations from a classical state.

\section{QCP of amplitude damping channel}

In this section, we will show exactly how to calculate the QCP by
providing an example. The amplitude damping (AD) channel
$\Lambda^{\mathrm{AD}}$ describes the evolution of a quantum system
interacting with a zero-temperature bath. The operator-sum
presentation of AD channel is
$\Lambda^{\mathrm{AD}}=\sum_{i=0}^1E_i^{\mathrm{AD}}(\cdot)E_i^{\mathrm{AD}\dagger}$,
where $E_0^{\mathrm{AD}}=|0\rangle\langle0|+\sqrt
p|1\rangle\langle1|$ and
$E_1^{\mathrm{AD}}=\sqrt{1-p}|0\rangle\langle1|$. Here we choose
quantum discord and one-way quantum deficit as measure of quantum
correlation in Eq. (\ref{QCP}). Now we are ready to calculate the
QCP of AD channel.

According to theorem 1, the optimal input state should be of form
\begin{equation}
\rho=q_1|\theta\rangle\langle\theta|\otimes|0\rangle\langle0|+q_2|\theta+\frac{\pi}{2}\rangle\langle\theta+\frac{\pi}{2}|\otimes|1\rangle\langle1|,\label{optimal_input_2}
\end{equation}
where $|\theta\rangle=\cos\theta|0\rangle+\sin\theta|1\rangle$.
Intuitively, $q_1=q_2=1/2$ should be chosen to maximize the initial
classical correlation, while $\theta=\pi/4$ should hold such that
the coherence between the two energy levels $|0\rangle$ and
$|1\rangle$ of qubit $A$ is maximized. These are verified by
numerical results.

Depending on the above discussion, the analytical expression of QCP
defined on quantum discord and one-way quantum deficit are
respectively
\begin{equation}
\mathcal
Q_{\delta}(\Lambda^{\mathrm{AD}})=h(p)+h(\sqrt{1-p})-h(\sqrt{1-p+p^2})-1,
\end{equation}
and
\begin{eqnarray}
\mathcal Q_{\Delta}(\Lambda^{\mathrm{AD}})=\min\{h(\sqrt{1-p})-h(\sqrt{1-p+p^2}),\nonumber\\
h(p)-h(\sqrt{1-p+p^2}),\nonumber\\
\frac{h(t_1)+h(t_2)}{2}-h(\sqrt{1-p+p^2})\},
\end{eqnarray}
where
$h(x)=-\frac{1+x}{2}\log_2\frac{1+x}{2}-\frac{1-x}{2}\log_2\frac{1-x}{2}$,
$t_1=\sqrt{1-p}\sin2\chi+p\cos2\chi$,
$t_2=\sqrt{1-p}\sin2\chi-p\cos2\chi$, where $\chi$ satisfies
\begin{equation}
\tan2\chi=\frac{\sqrt{1-p}\log_2\frac{(1+\sqrt{1-p}\sin2\chi)^2-(p\cos2\chi)^2}{(1-\sqrt{1-p}\sin2\chi)^2-(p\cos2\chi)^2}}
{p\log_2\frac{(1+p\cos2\chi)^2-(\sqrt{1-p}\sin2\chi)^2}{(1-p\cos2\chi)^2-(\sqrt{1-p}\sin2\chi)^2}}.
\end{equation}

\begin{figure}
\scalebox{0.4}[0.4]{\includegraphics{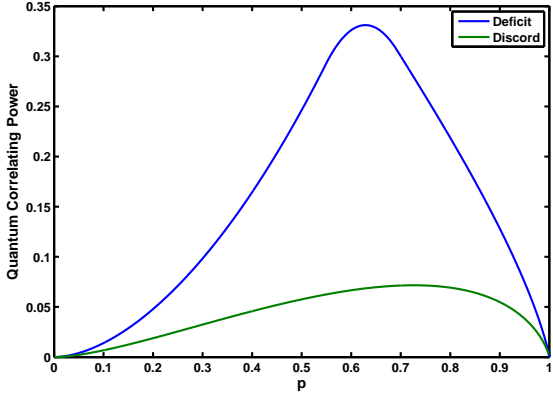}}
\caption{\label{fig1} Quantum-correlating power of amplitude damping
channel against the parameter $p$ of AD channel. Blue and green
lines are respectively the QCP based on one-way quantum deficit and
quantum discord.}
\end{figure}

\begin{figure}
\scalebox{0.4}[0.4]{\includegraphics{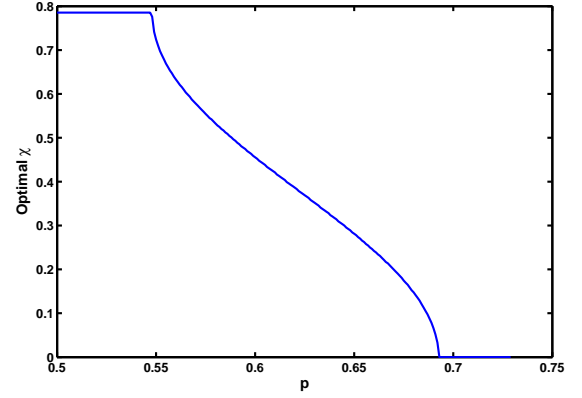}}
\caption{\label{fig2} Optimal measurement basis
$\{|\chi\rangle,|\chi+\pi/2\rangle\}$ for one-way quantum deficit
against the parameter $p$ of AD channel, where
$|\chi\rangle=\cos\chi|0\rangle+\sin\chi|1\rangle$.}
\end{figure}

The optimal measurement basis $\{|\chi\rangle,|\chi+\pi/2\rangle\}$
in the definition of one-way quantum deficit as in Eq.
(\ref{deficit}) transfers gradually from $\{|+\rangle,|-\rangle\}$
to $\{|0\rangle,|1\rangle\}$, as shown in Fig. \ref{fig2}, while for
quantum discord, the optimal measurement is always
$\{|+\rangle,|-\rangle\}$.

\section{Super-activation of QCP}

In this section, we will claim an interesting property of QCP.
Consider two classical-quantum states $\rho_{AB}$ and $\rho_{A'B'}$
with qubits $A$ and $A'$ at one site and qubits $B$ and $B'$ at
another. A local two-qubit unitary operator acting on qubits $A$ and
$A'$ can activate two zero-QCP single-qubit channels into a positive
QCP two-qubit channel. We call this phenomenon the super-activation
of QCP.

We here give an example of phase-damping (PD) channel to show
exactly how this property works. The Kraus operators of PD channel
are
$E_0^{\mathrm{PD}}=|0\rangle\langle0|+\sqrt{1-p}|1\rangle\langle1|$
and $E_1^{\mathrm{PD}}=\sqrt{p}|1\rangle\langle1|$. Clearly, PD
channel is a mixing channel, which means that quantum correlation
cannot be created when a single copy of classical-quantum state is
considered.

Now consider initial state of qubits $A$ and $B$
\begin{equation}
\rho_{AB}=\frac12\sum_{i=0}^1|i\rangle_A\langle
i|\otimes|i\rangle_B\langle i|.
\end{equation}
Qubits $A'$ and $B'$ are in the same state, then the total state of
the four qubits is
\begin{eqnarray}
\rho&=&\frac14\rho_{AB}\otimes\rho_{A'B'}\nonumber\\
&=&\frac14\sum_{i,j}|ij\rangle_{AA'}\langle
ij|\otimes|ij\rangle_{BB'}\langle ij|.
\end{eqnarray}
Now apply a two-qubit unitary operation $U$:
$U|ij\rangle=|\psi_{ij}\rangle$ on qubits $A$ and $A'$, where
$|\psi_{00}\rangle=\frac{1}{\sqrt2}(|00\rangle+|11\rangle)$,
$|\psi_{11}\rangle=\frac{1}{\sqrt2}(|0+\rangle+|1-\rangle)$,
$|\psi_{01}\rangle=\frac{1}{\sqrt2}(|01\rangle-|10\rangle)$, and
$|\psi_{10}\rangle=\frac{1}{\sqrt2}(|0-\rangle-|1+\rangle)$. Then
qubits $A$ and $A'$ each transmits through a PD channel, and the
output state becomes $\rho'=
\Lambda_A^{\mathrm{PD}}\otimes\Lambda_{A'}^{\mathrm{PD}}\otimes
I_{BB'}(U_{AA'}\rho U_{AA'}^{\dagger})$. Now we check whether
quantum correlation defined on $AA'$ is created between the
bipartition $AA':BB'$ by using the criterion in Ref.
\cite{PhysRevA.85.032102}. Notice that
\begin{eqnarray}
&&[\Lambda^{\mathrm{PD}}\otimes\Lambda^{\mathrm{PD}}(\psi_{00}),
\Lambda^{\mathrm{PD}}\otimes\Lambda^{\mathrm{PD}}(\psi_{11})]\nonumber\\
&=&\frac18\tilde ip\sqrt{1-p}(I\otimes\sigma_y+\sigma_y\otimes
I)\neq0,
\end{eqnarray}
and consequently, quantum correlation is created between the
bipartition $AA':BB'$.

The super-activation of QCP is a collective effect. The reduced
two-qubit states
$\rho'_{AB}=\mathrm{Tr}_{A'B'}(\rho')=(I_A/2)\otimes\rho_B$ and
$\rho'_{A'B'}=\mathrm{Tr}_{AB}(\rho')=(I_{A'}/2)\otimes\rho_{B'}$
are product states, which contain no correlations at all. The local
two-qubit unitary operation $U$ does not build correlations between
qubits $A$ and $A'$, since reduced state of qubits $A$ and $A'$
remains completely mixed during the whole process. All in all, no
correlation exists between any two qubits of the four-qubit state
$\rho'$. Therefore, we suppose that the effect of super-activation
of QCP is due to the genuine quantum correlation.

\section{Conclusion}

We have introduced the concept of quantum-correlating power for
quantifying the ability of local quantum channel to generate quantum
correlation from a classically correlated state. For any channel,
the general form of the optimal input state has been proved to be
the classical-classical state. Furthermore, the single-qubit
channels with maximum QCP can be found in the class of local
channels which takes a classical-classical state to a rank-2
quantum-classical states. The explicit expression for QCP of
single-qubit AD channel has been obtained.

When two zero-QCP channels are used together, a positive-QCP channel
can be obtained. We call this effect the super-activation of QCP. In
the example of PD channel, we find a four-qubit state with genuine
four-qubit quantum correlation but zero two-qubit correlation. This
result should be helpful in the study of quantum correlating
structure in multi-qubit states.

\begin{acknowledgments}
This work was supported by the NKBRSFC under grants Nos.
2012CB922104, 2011CB921502, 2012CB821305, 2009CB930701,
2010CB922904, NSFC under grants Nos. 10934010, 60978019, 10975181,
11175247 and NSFC-RGC under grants Nos. 11061160490 and
1386-N-HKU748/10.

\end{acknowledgments}

%\newpage %Just because of unusual number of tables stacked at end

\begin{thebibliography}{29}
\bibitem[{foo()}]{footnote}
{We consider AD channel with $p^2<1-p$ acting on qubit $A$ of
  the input state
  $\rho=(|+0\rangle_{AB}\langle+0|+|-\phi\rangle_{AB}\langle-\phi|)/2$, where
  $|\phi\rangle$ is a pure state satisfying
  $\langle\phi|\sigma_3|\phi\rangle=-[3p^2-(1-p)]/[p^2+(1-p)]$. the geometric
  quantum discord that can be created by the channel is
  $Q_G^{max}=(p^2(1-p))/(2[p^2+(1-p)])$. However, when we consider input
  state with $|\phi\rangle=|1\rangle$, the corresponding geometric quantum
  discord that can be created is $Q_G=p^2<Q_G^{max}$. It means that local
  operation on $B$ with Kraus operators $E_0=|0\rangle\langle0|$ and
  $E_1=|\phi\rangle\langle1|$ can increase geometric discord on $A$.}
\end{thebibliography}

\end{document}